\documentclass[journal=nalefd,manuscript=letter]{achemso}

\usepackage[T1]{fontenc} 
\usepackage{graphicx} 
\usepackage{amsbsy} 
\usepackage{amssymb,amsmath}
\usepackage{braket}
\usepackage{color}
\usepackage{ulem}
\usepackage{mathrsfs}

\author{Seongphill Moon}
\affiliation{National High Magnetic Field Laboratory, Tallahassee, Florida 32310, USA}
\alsoaffiliation{Department of Physics, Florida State University, Tallahassee, Florida 32310, USA}
\author{Yuxuan Jiang}
\affiliation{School of Physics and Optoelectronic engineering, Anhui University, Hefei, Anhui 230601, China}
\alsoaffiliation{Center of Free Electron Laser and High Magnetic Field, Anhui University, Hefei, Anhui 230601, China}
\email{yuxuan.jiang@ahu.edu.cn}
\author{Jennifer Neu}
\affiliation{National High Magnetic Field Laboratory, Tallahassee, Florida 32310, USA}
\alsoaffiliation{Department of Physics, Florida State University, Tallahassee, Florida 32310, USA}
\author{Theo Siegrist}
\affiliation{National High Magnetic Field Laboratory, Tallahassee, Florida 32310, USA}
\alsoaffiliation{Department of Chemical and Biomedical Engineering, FAMU-FSU College of Engineering, Tallahassee, FL, 32310, USA}
\author{Mykhaylo Ozerov}
\affiliation{National High Magnetic Field Laboratory, Tallahassee, Florida 32310, USA}
\author{Zhigang Jiang}
\affiliation{School of Physics, Georgia Institute of Technology, Atlanta, Georgia 30332, USA}
\email{zhigang.jiang@physics.gatech.edu}
\author{Dmitry Smirnov}
\affiliation{National High Magnetic Field Laboratory, Tallahassee, Florida 32310, USA}
\email{smirnov@magnet.fsu.edu}

\title{Magneto-optical evidence of tilting effect in coupled Weyl bands} 
\begin{document}
\begin{tocentry}
\centerline{\includegraphics[width=0.57\textwidth]{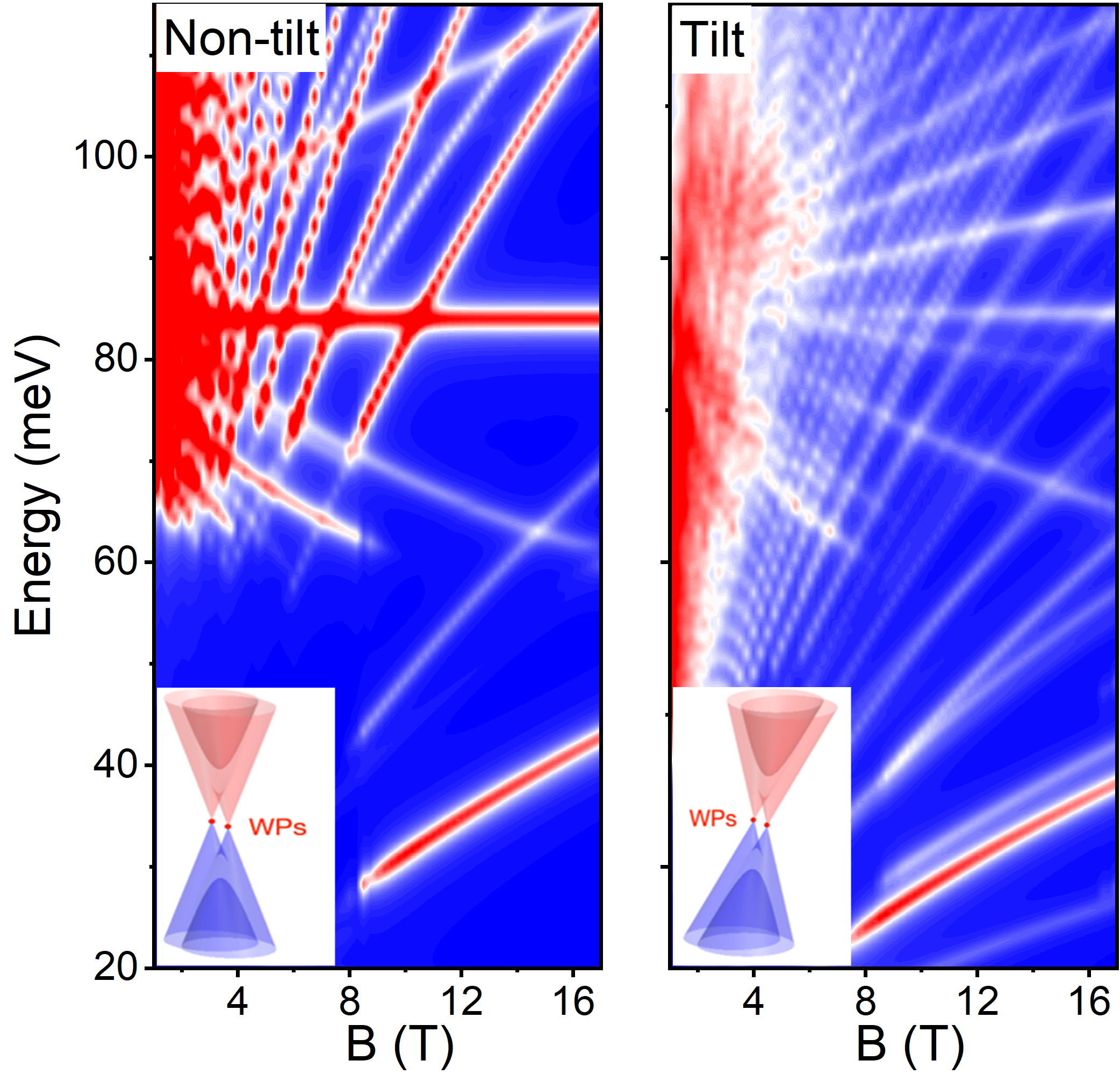}}
\end{tocentry}

\begin{abstract}
Theories have revealed the universality of the 
band tilting effect in topological Weyl semimetals (WSMs) and its implications for the material's physical properties. However, the experimental identification of tilted Weyl bands remains much less explored. Here, by combining magneto-infrared optical studies with a four-band coupled Weyl point model, we report spectroscopic evidence of the tilting effect in the well-established WSM niobium phosphide. Specifically, we observe Landau level transitions with rich features that are well reproduced within a model of coupled tilted Weyl points.
Our analysis indicates that the tilting effect relaxes the selection rules and gives rise to  transitions that would otherwise be forbidden in the non-tilt case. Additionally, we observe unconventional interband transitions with flat and negative magnetic field dispersions, highlighting the importance of coupling between Weyl points. Our results not only emphasize the significance of the tilting effect in the optical responses of WSMs but also demonstrate magneto-optics as an effective tool for probing the tilting effect in electronic band structures.
\end{abstract}

\section{Maintext}
Weyl semimetals (WSMs) represent a compelling class of topological quantum matter where the conduction band (CB) and valence band (VB) intersect at discrete points, known as Weyl points (WPs). These points host a linear band dispersion in their vicinity with specific chiralities, resembling Weyl particles in high-energy physics.\cite{Vafek_review_2014,Jia_review_2016,Yan_review_2017,RMP_WeylDirac} These unique band structures give rise to many exotic properties, such as Fermi arc surface states,\cite{Fermi_arc_Prediction,Fermi_arc_Ran_2011,Fermi_arc_Balents_2011,TaASprediction1,TaAsprediction2,Fermiarccalculation} chiral anomaly,\cite{chiralanomalymechanism} negative magnetoresistance,\cite{NMRTaAs1,NMRTaAs2,NbPNMR} giant second harmonic generation,\cite{SHG} and a colossal photovoltaic effect.\cite{PhotocurrentTiltExp,ColossalPhotovoltaic}

Since WPs are typically located at low-symmetry points in the Brillouin zone (BZ), their band structures are often accompanied by tilting.\cite{Weyl_type_Bernevig_2015,NbPtypeII,NbPBand} The tilting effect is considered a solid-state realization of Lorentz violation,\cite{xu2017discovery,yan2017lorentz} and it can significantly alter the transport and optical properties of WSMs, leading to intriguing phenomena such as chiral photocurrents,\cite{PhotocurrentTiltExp,PhotocurrentTiltTheory} the anisotropic chiral magnetic effect,\cite{van2017anisotropic} and unconventional optical selection rules \cite{Weyltiltingeffect,YuxuanTiltTheory,wyzula2022lorentz}. Even though tilting can be probed by angle-resolved photoemission spectroscopy,\cite{xu2017discovery,jiang2017signature,yan2017lorentz} its influence on the material's physical properties remains much less explored experimentally.

\begin{figure*}[t!]
   \centering
\includegraphics[width=\linewidth]{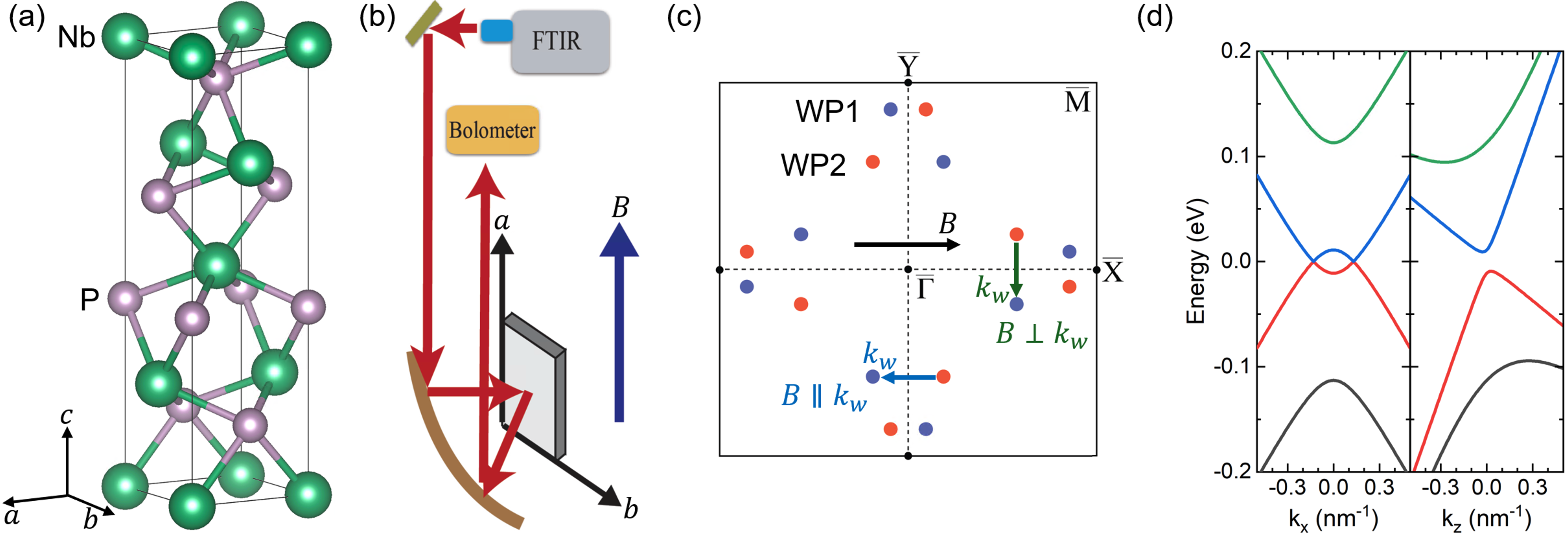}
    \caption{(a) Unit cell of the tetragonal crystal lattice of NbP. (b) Schematic of the experimental setup for IR Voigt reflection measurements.  
    (c) Schematic of the (001) surface BZ with projected bulk WP1 and WP2. The black arrow shows the direction of the applied magnetic field. The red and blue dots represent the WPs with different chirality, and vector $k_w$ describes the orientation and separation of two WPs in the same pair. (d) Zero-field band structure of tilted WP2 in selected directions.}
    \label{fig1}
\end{figure*}

The transition-metal monopnictide family (TaAs, TaP, NbAs, NbP) includes the first experimentally observed WSMs,\cite{TaAsAPRES1,TaAsARPES2,NbPTaPARPES,NbPARPES,NaAsARPES,TaPARPES} which are also expected to exhibit tilted WPs.\cite{YuxuanTiltTheory,NbPBand,NbPBand_2} Later magneto-infrared (magneto-IR) optical measurements reveal rich spectral features in these materials that cannot be simply explained by an isolated WP model,\cite{NbPYuxuanFaraday,NbAs_chiralLL,polatkan2020magneto,lu2022weyl,zhao2022unconventional} and requires the consideration of more realistic structures such as coupled WPs\cite{NbPYuxuanFaraday,zhao2022unconventional} or partially gapped nodal loops.\cite{polatkan2020magneto} However, the manifestation of tilting effect was not studied in these materials. In this work, we perform IR magneto-reflection spectroscopy on NbP in the Voigt geometry and compare the experimental results to the coupled WP model, both with and without the tilting effect. We find that the non-tilt model cannot adequately describe the rich structure of spectral features observed in the experiment, leaving many low-energy transitions unexplained. The tilting of coupled WPs relaxes the optical selection rules and allows transitions that are forbidden in the non-tilt case. These forbidden transitions serve as spectroscopic evidence of tilted WPs in WSMs. Our work highlights the necessity of considering the tilting effect to fully understand the physical properties of WSMs.

The NbP single crystal studied here was grown using the chemical vapor transport method (see Supporting Information), and it crystallized into a body-centered tetragonal lattice structure with space group $I4_1md$ (No. 109) and point group C$_{4v}$. The crystal structure of NbP is shown in Figure \ref{fig1}a, where the crystallographic $a$, $b$, and $c$ axes correspond to the $k_x$, $k_y$, and $k_z$ axes, respectively, in the following discussion of band dispersion.

The lattice structure of NbP exhibits time-reversal symmetry but lacks inversion symmetry, a key requirement for the formation of (non-magnetic) WPs. NbP hosts two types of WPs located at different $k_z$ planes. We designate the WPs at $k_z=0$ as WP1, while those away from the $k_z=0$ plane are referred to as WP2. Due to the mirror and fourfold rotational symmetries in NbP, there are 4 pairs of WP1 and 8 pairs of WP2 in the first BZ. Half of the WPs are oriented along the $k_x$ axis, while the other half are along the $k_y$ axis. Figure \ref{fig1}c illustrates the distribution and orientation of the WPs in the (001) surface plane projection. The orientation of each WP pair is indicated by a $k_{w}$ vector, connecting WPs of opposite chirality.

The band structure of the WPs in NbP can be well described by a four-band tilted coupled WP model with a $4\times 4$ Hamiltonian\cite{koshino1,YuxuanTiltTheory}
\begin{equation}
H=v\tau_x(\boldsymbol{\sigma}\cdot\mathbf{p})+ m\tau_z+b\sigma_x+T(\mathbf{p}),
\label{Ham_1}
\end{equation}
where $\mathbf{p}=(p_x,p_y,p_z)$ is the crystal momentum, and $\sigma=(\sigma_x,\sigma_y,\sigma_z)$ and $\tau=(\tau_x,\tau_y,\tau_z)$ represent the spin and orbital Pauli matrices, respectively. The Fermi velocity $v$, hybridization gap $m$, and intrinsic Zeeman effect $b$ are material-specific band parameters. The last term, $T(\mathbf{p})=v(t_xp_x\tau_x+t_yp_y+t_zp_z)$, describes the tilting effect of the band structure, and it is parameterized by $\mathbf{t}=(t_x,t_y,t_z)$. Here, $|t|<1$ ($|t|>1$) corresponds to the type I (type II) tilted WPs, respectively. In this model, the intrinsic Zeeman effect leads to the formation of WPs, and the hybridization gap $m$ arises from the coupling between WPs,\cite{koshino1,NbPYuxuanFaraday} resulting in four bands at the $\Gamma$-point of the BZ (i.e., the upper CB, lower CB, lower VB, and upper VB, in decreasing energy order, as illustrated in Figure 1d).
Near the WPs, this model shows excellent agreement with results from {\it ab initio} calculations across a broad range of momentum and energy,\cite{YuxuanTiltTheory} providing an accurate description of the magneto-IR experiments \cite{NbPYuxuanFaraday,zhao2022unconventional}. 

To reveal the tilting effect, we perform IR magneto-reflection measurements on NbP in the Voigt geometry. A schematic of the experimental setup is shown in Figure \ref{fig1}b. In our measurements, the magnetic field lies along the $a$ axis, while the light propagation direction is perpendicular to the $ab$ plane. Based on the band structure from first-principles calculations \cite{NbPBand, NbPBand_2, NbPtypeII}, the Voigt geometry measurements offer two major advantages. First, WP1 in NbP has zero Fermi velocity along the $z$ axis,\cite{YuxuanTiltTheory,NbPBand} leading to a very flat band dispersion. This suggests that the cyclotron orbit is infinitely large, allowing us to neglect WP1 Landau levels (LLs) when analyzing the magneto-IR spectra. Second, the tilting effect is stronger in WP2 in NbP. WP1 tilts along the $y$ direction ($t_y=0.4$), while WP2 tilts primarily along the $k_z$ direction, with a small component along $k_y$ ($t_y=0.1$ and $t_z=0.55$).\cite{YuxuanTiltTheory} Figure \ref{fig1}d shows the tilted band dispersion of WP2 along the $k_x$ (with $k_y=0,\ k_z=0$) and along the $k_z$ (with $k_x=0,\ k_y=0$) directions using the band parameters extracted from our experiment.

\begin{figure*}[t!]
   \centering
\includegraphics[width=0.8\linewidth]
{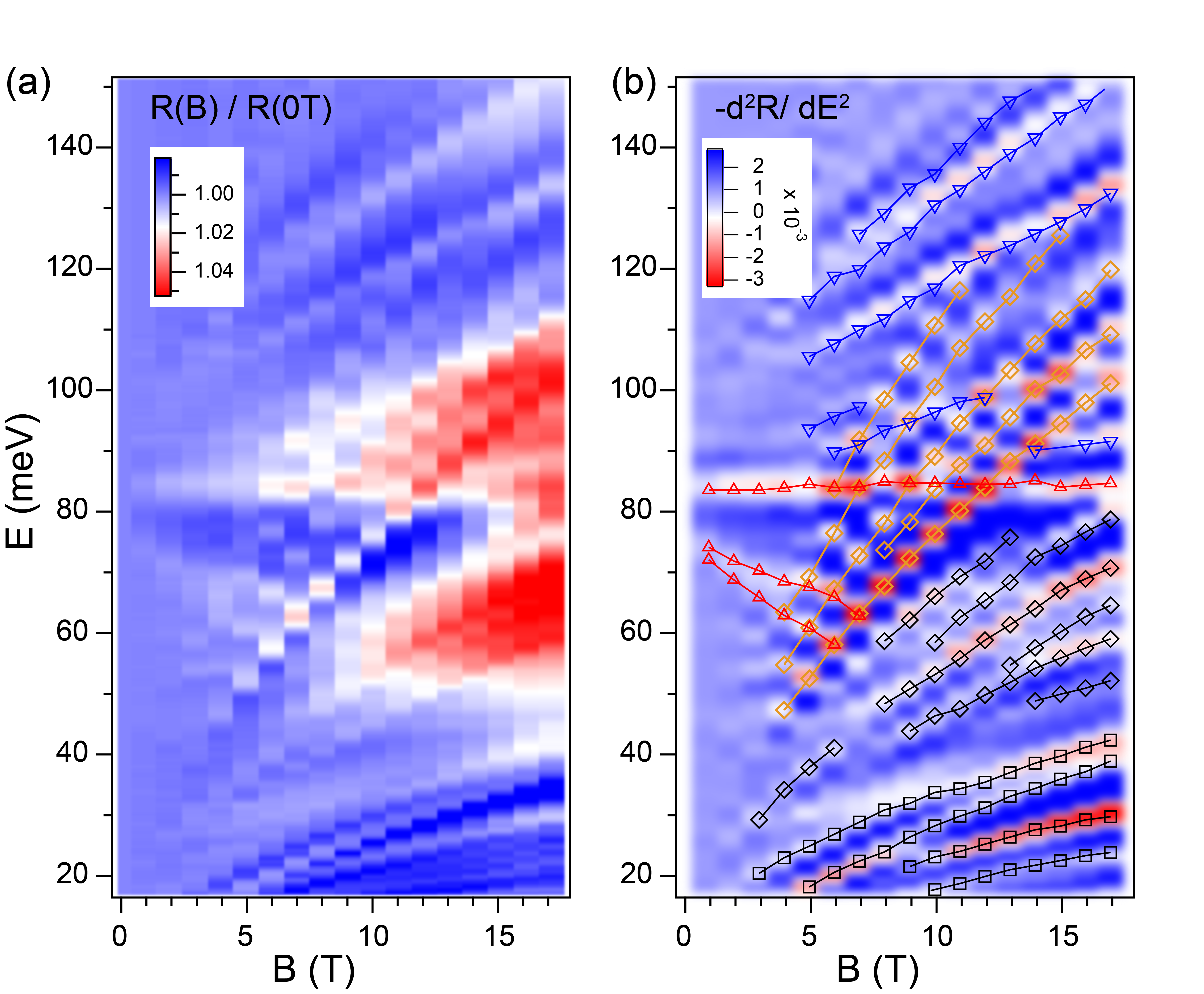}
    \caption{Magneto-reflection measurement results of NbP.  False-color plot of the normalized magneto-reflectance spectra, $R(B)/R(B=0\text{T})$, measured at 5K (a), and its second derivative, $d^2R/dE^2$ (b). Symbols are extracted transition energies grouped into four sets and color-coded accordingly.
    }
    \label{fig2}
\end{figure*}

Figure \ref{fig2}a shows the false-color plot of the magneto-reflectance spectra of NbP, normalized by the zero-field spectrum, $R(B)/R(B=0\text{T})$, from 1 T to 17 T. Normalized spectra at each magnetic field can be found in Supporting Information. As the magnetic field increases, we observe multiple series of LL transitions, which exhibits distinct energy intercepts or magnetic field dispersions. For better contrast and clear visibility of the modes, we take the second derivative of the normalized magneto-reflectance spectra to energy and plot it in Figure \ref{fig2}b. Figure \ref{fig2}b also displays the extracted transition energies of the observed modes at different magnetic fields, and we divide them into four different groups and color-code them accordingly based on our detailed comparison with calculations below. Nevertheless, without detailed analysis, we can already distinguish between interband- and intraband-like transitions by their zero magnetic field intercept. 
 
For the sets of LL transitions plotted in black and orange, the zero-field intercept approaches zero, and they show different magnetic field dependence at finite fields. Conversely, for the red and blue sets of LL transitions, the zero-field intercept is around 84 meV, indicating an interband origin. Among these transitions, the red set (Figure \ref{fig2}b) is particularly noteworthy. First, it includes a nearly flat transition in magnetic field at 84 meV. We rule out the possibility of an IR-active phonon mode, as no such mode has been observed or predicted near this energy \cite{opticalconductivityNbP}. Second, we observe a faint but visible transition with negative magnetic field dispersion, starting from a similar zero-field intercept. Such unconventional LL transitions, with both flat and negative magnetic field dispersions, have also been reported in TaP and NbAs within the same WSM family.\cite{polatkan2020magneto}

\begin{figure}[t!]
   \centering
\includegraphics[width=0.8\linewidth]{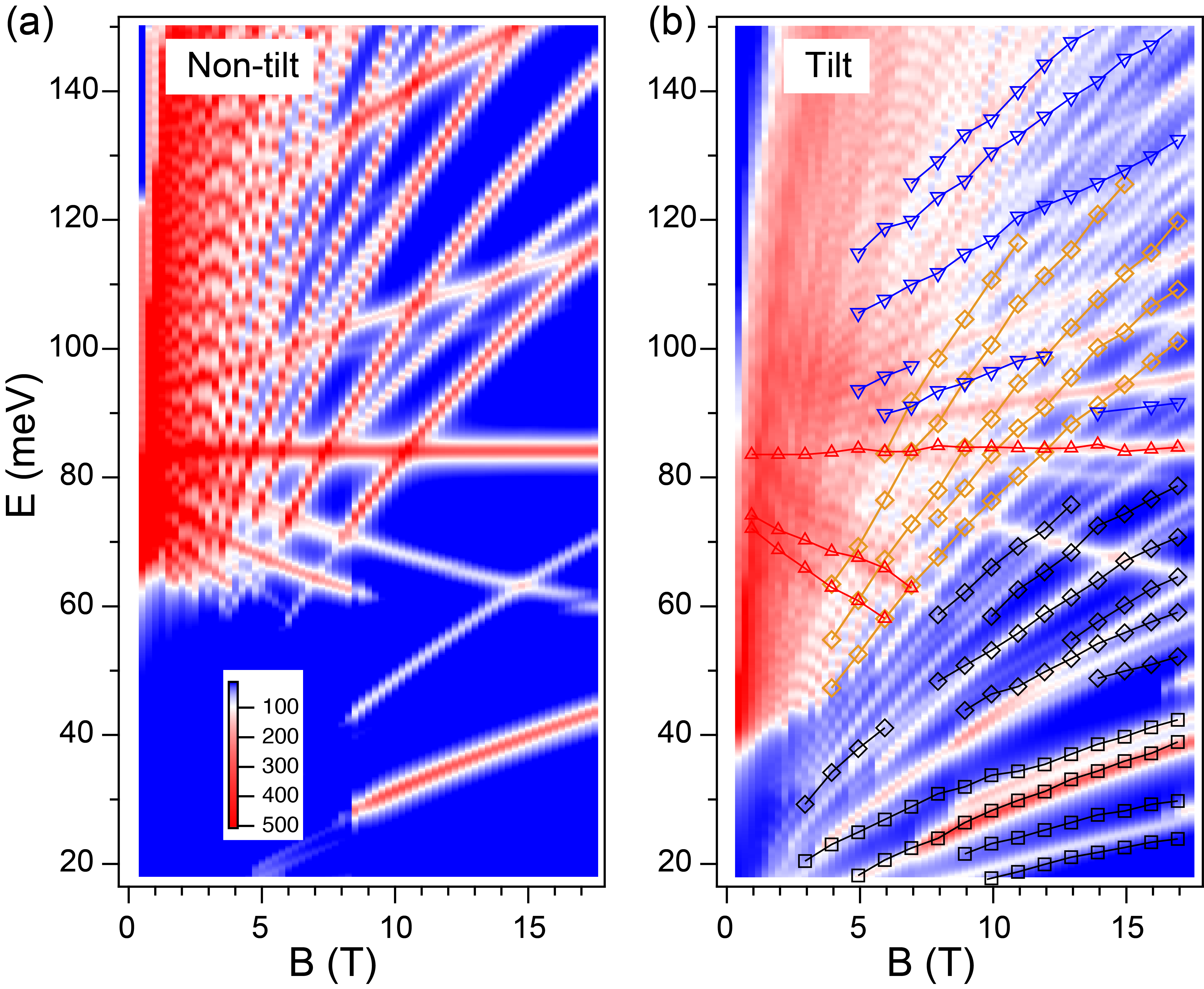}
    \caption{Calculated magneto-absorption spectra of NbP from (a) a non-tilt model and (b) a tilt model. The symbols are inter-LL transition energies plotted in Figure \ref{fig2}b.}
    \label{fig3}
\end{figure}

To analyze the magneto-reflectance spectra of NbP, we calculate the LLs using the WP Hamiltonian 
(Eq. \ref{Ham_1})
with Peierls substitution. The optical transitions between LLs are then computed using Fermi's golden rule.
We consider only the optical weight from the $\Gamma$ point where the joint density of states diverges as can be seen from Figure \ref{fig1}d. More details of the calculations can be found in Supporting Information. The resulting magneto-absorption spectra are shown in Figures \ref{fig3}a and \ref{fig3}b, with or without the tilting effect, respectively. In the non-tilt case, we directly fit the experiment data and find that $b = 50$ meV, $m = 42$ meV, and $v= 3.3 \times 10^5$ m/s yield the best fit. In the tilt case, we set the tilt parameter to be $\mathbf{t}=(0,0.1,0.55)$ by fitting the four-band coupled model to the \textit{ab-initio} calculations\cite{YuxuanTiltTheory} and find that the experimental data are best fit with $b = 60$ meV, $m = 51$ meV, and $v= 4.1 \times 10^5$ m/s. The Pauli blocking effect in optical transitions is also accounted for by considering the Fermi level evolution in magnetic fields. The carrier density is set to 6 $\times$ 10$^{23}$ m$^{-3}$.

The calculated LL transitions for both the tilt and non-tilt cases exhibit similar structures to those observed in experiment. Both cases display interband transitions with flat, upward (positive slope), and downward (negative slope) dispersions in magnetic fields, along with two series of transitions that have zero-energy intercepts. The key difference between the two cases lies in the appearance of many
additional modes for the tilt case, indicating the relaxation of selection rules. Indeed, the tilting effect mixes the wavefunctions of different LLs, leading to the breaking of conventional selection rules and the redistribution of the optical weights.\cite{YuxuanTiltTheory,wyzula2022lorentz,Weyltiltingeffect,koshino2} 
It is obvious that the non-tilt model cannot reproduce that large number of inter-LL transitions and, more importantly, cannot explain the existence of several intense low-energy modes (below 60 meV).
In contrast, the tilt case calculation reproduces well the majority of inter-LL transitions, including the low-energy ones. Thus, we can conclude that the tilting effect plays a crucial role in the optical responses of NbP.

Next, we discuss the LL transitions in detail. In our experiment, the magnetic field direction is parallel to $k_{w}$ for half of the WPs and perpendicular to $k_{w}$ for the other half (as shown in Figure \ref{fig1}c). This configuration results in two different types of Landau quantizations, (i) $B\parallel k_{w}$ and (ii) $B\perp k_{w}$, requiring us to examine their associated LL transitions separately. 

\begin{figure*}[t!]
\includegraphics[width=\linewidth]{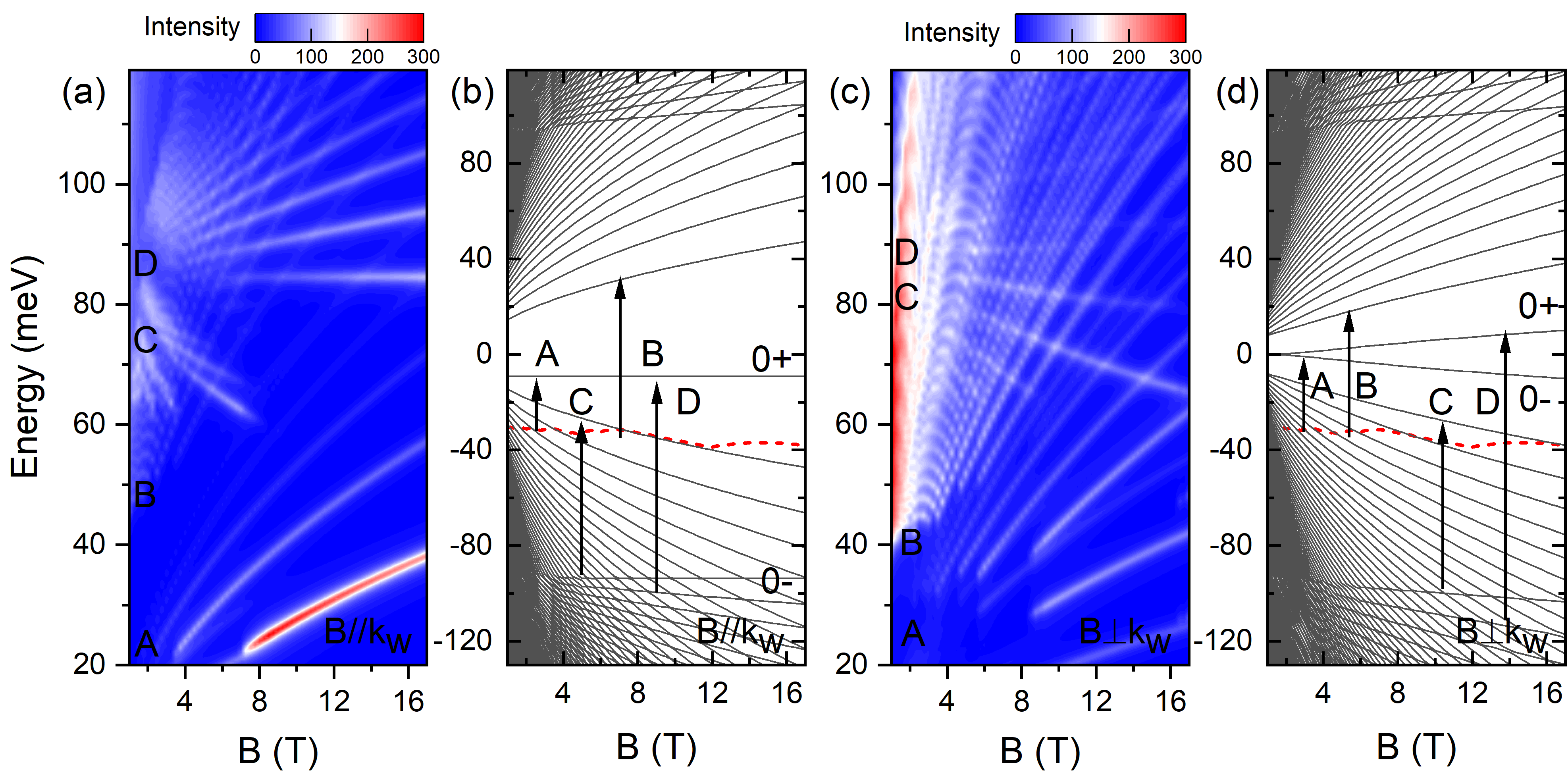}
    \caption{(a,c) False-color plots of the calculated magneto-absorption spectra in (a) $B\parallel k_{w}$ and (c) $B\perp k_{w}$ configurations. (b,d) The corresponding LLs calculated for (b) $B\parallel$ $k_{w}$ and (d) $B\perp$ $k_{w}$. The red dash lines represent a shared Fermi level between these two configurations with a hole carrier density of 6 $\times$ 10$^{26}$ m$^{-3}$. Letters A, B, C, and D label four distinct sets of LL transitions, color-coded in black,  yellow, red, and blue, respectively, in Figure \ref{fig2}b. Their representative transitions are indicated by the arrows in Figure \ref{fig2}b and d.}
    \label{fig4}
\end{figure*}

Figure \ref{fig4} shows the calculated LL transitions and the resulting magneto-absorption spectra for both $B\parallel k_{w}$ and $B\perp k_{w}$. In both cases, the LL transitions are grouped into four distinct series, labeled by the Roman numerals A, B, C, and D, respectively, and the associated arrow indicates representative transitions. For $B\parallel k_{w}$, the downward and flat interband transitions (C-series) originate from the $0-$ LL in the upper VB to the LLs in the lower VB. The upward interband transitions (D-series) occur between the LLs in the upper VB and the 0+ LLs in the lower VB. The B-series transitions are interband transitions between the lower VB and CB. The A-series transitions originate from intraband transitions within the lower VB. For $B\perp k_{w}$, the LL transition assignments are similar. 

Based on the above analysis, we attribute the red and blue transitions in experiment (Figure \ref{fig2}b) to the C- and D-series in calculation, and the dominant contributions are from the $B\parallel k_{w}$ case. These interband transitions, especially the flat and downward modes, signify the importance of the four-band coupled WP model in understanding the optical transitions in WPs, where the upper VB and CB exhibit different magnetic field dispersion from the lower VB and CB \cite{koshino1}. In contrast, two-band models feature only one dispersionless 0th LL, and all interband and intraband LL transitions exhibit positive magnetic field dispersion \cite{zhang2016linear,lu2022weyl,polatkan2020magneto}. Hence, a two-band model cannot account for the experimentally observed transitions with flat and downward dispersions.

For the black and orange transitions observed in experiment (Figure \ref{fig2}b), we attribute these to the A- and B-series in calculation. However, due to the gap between the lower VB and CB in the B $\parallel$ $k_{w}$ case (Figure \ref{fig4}b), we find that the majority of these transitions occur in the $B\perp k_{w}$ case, as shown in Figures \ref{fig4}c and \ref{fig4}d.

In conclusion, we performed the experimental investigation of the magneto-IR reflectance of WSM NbP in the Voigt geometry and found a complex structure of inter-LL transitions. We observe that the interband LL transitions exhibit not only the conventional positive magnetic field dispersion but also flat and negative dispersions. Through theoretical analysis using a coupled WP model, we attribute these unusual dispersions to interband transitions between the upper and lower VBs, which cannot be explained by an isolated WP model. Furthermore, we demonstrate that most of the strong low-energy transitions observed in experiment are forbidden in the no-tilt model but can be semi-quantitatively explained by introducing the tilting effect in the model analysis. These forbidden transitions serve as spectroscopic evidence of tilted Weyl bands and signify the importance of the tilting effect in understanding the band structure and optical properties of WSMs, as it redistributes wavefunctions and breaks conventional selection rules.


\begin{acknowledgement}
This work was primarily supported by the Department of Energy (Grant No. DE-FG02-07ER46451). Y.J. acknowledges support from the National Natural Science Foundation of China (Grant No. 12274001), the Natural Science Foundation of Anhui Province (Grant No. 2208085MA09), and the Department of Education of Anhui Province (Grant No. 2023AH020004). The magneto-IR measurements were performed at the National High Magnetic Field Laboratory, which is supported by the National Science Foundation Cooperative Agreement (No. DMR-2128556) and the State of Florida. 
\end{acknowledgement}
\begin{suppinfo}
See Supporting Information for details on sample growth, model calculations, and normalized spectra from magneto-reflectance measurements.
\end{suppinfo}

\bibliography{NbP_bib}

\end{document}